\documentclass[%
 aip,
rsi,%
 amsmath,amssymb,  
 reprint,%
]{revtex4-1}

\usepackage{graphicx}
\usepackage{dcolumn}
\usepackage{bm}
\usepackage{graphicx}
\usepackage{color}
\usepackage{bm}
\usepackage{ulem}
\usepackage{soul} 
\begin{document}

\preprint{AIP/123-QED}


\title{Influence of symmetry breaking on  Fano-like resonances in high Figure of Merit planar terahertz metafilms}



\author{Joshua A. Burrow}
\email{burrowj2@udayton.edu}
\affiliation{Electro-Optics Department, University of Dayton, 300 College Park Ave., Dayton, OH, 45469, USA}

\author{Riad Yahiaoui}%
\affiliation{Department of Physics $\&$ Astronomy, Howard University, 2355 6th St. NW, Washington, DC, 20059, USA}%

\author{Wesley Sims}%
\affiliation{Department of Physics, Morehouse College, 830 Westview Dr. SW, Atlanta, GA 30314, USA}%

\author{Zizwe Chase}%
\affiliation{Department of Physics $\&$ Astronomy, Howard University, 2355 6th St. NW, Washington, DC, 20059, USA}%

\author{Viet Tran}
\affiliation{Department of Physics $\&$ Astronomy, Howard University, 2355 6th St. NW, Washington, DC, 20059, USA}%

\author{Andrew Sarangan}
\affiliation{Electro-Optics Department, University of Dayton, 300 College Park Ave., Dayton, OH, 45469, USA}%

\author{Jay Mathews}
\affiliation{Physics Department, University of Dayton, 300 College Park Ave., Dayton, OH, 45469, USA}%

\author{Willie S. Rockward}%
\affiliation{Department of Physics, Morgan State University, 1700 E. Cold Spring Ln., Baltimore, MD 21251, USA}%

\author{Imad Agha}
\affiliation{Electro-Optics Department, University of Dayton, 300 College Park Ave., Dayton, OH, 45469, USA}%
\affiliation{Physics Department, University of Dayton, 300 College Park Ave., Dayton, OH, 45469, USA}%

\author{Thomas A. Searles}
\email{thomas.searles@howard.edu}
\affiliation{Department of Physics $\&$ Astronomy, Howard University, 2355 6th St. NW, Washington, DC, 20059, USA}

\date{\today}

\begin{abstract}
It is well established that nearly all high-quality (Q) Fano-like resonances in terahertz (THz) metasurfaces broaden as asymmetry increases, resulting in a decline of Q-factor and an increase in the resonance intensity. Therefore, in order to determine the optimal design for applications in THz sensing, a Figure of Merit (FoM) is required.  Previous studies have identified the asymmetry regimes at which the peak FoM occurs for various, specific unit cell geometries. However to date, there is no systematic comparison of the resulting FoMs for common and novel geometries.  Here, a THz planar metafilm featuring split ring resonators with four distributed capacitive gaps is investigated to compare three unique methods of implementing asymmetry: (1) adjacent L-bracket translation, (2) capacitive gap translation  and (3) increasing gap width.  The results obtained find that by translating two gaps and increasing the bottom gap width of the unit cell, the high-Q Fano-like resonances are $6 \times$ higher than the FoM for the fundamental dipole mode.  This work further informs the design process for THz metasurfaces and as such will help to define their applications in photonics and sensing.
\end{abstract}


\maketitle 

\section{Introduction}
Metamaterials are sub-wavelength periodic unit cells engineered to exhibit effective macroscopic optical properties that can be exploited for the alteration of electromagnetic (EM) waves. 
The key attributes of metamaterials are (1) their properties are governed by their constituent material characteristics and geometry - not by chemical composition, and (2) their scalable EM response governed by the size of the sub-wavelength structures. 
The seminal unit cell to be demonstrated as a suitable geometry for metamaterials is the split ring resonator (SRR), introduced at the turn of the century to exhibit extraordinary spectral features unachievable by natural materials\cite{Pendry99,Smith00,Shelby01}.

Since the experimental verification of the SSR as a ``building block" for metamaterials, many unit cell designs have been explored by implementing subtle structural variations into the SRR such as weak asymmetry \cite{Singh:11}, relative super-cell adjustments \cite{Al-Naib:15,Zhang:12supercell}, and nested resonators\cite{Hussain2012NestedMMs,Wang2016Nested}.  
One such variation, the symmetric 4-gap square SSR geometries, exhibits polarization insensitive unit cells ideal for sensing applications with previous reports in the optical\cite{Chen15}, infrared\cite{Chen11}, microwave\cite{Penciu08} and THz\cite{Gong16,Wang16} regimes.

Breaking the symmetry in the unit cells of planar metamaterials enables access to different resonant modes, which cannot be excited in the symmetric configuration \cite{Singh:10}. These resonant modes have been termed as dark, sub-radiant or trapped because they couple weakly to free space and require an external perturbation, such as symmetry-breaking of the resonator geometry, in order to be excited \cite{Zhange1501142,Yang17,PhysRevLett.99.147401,PhysRevB.79.085111Singhbrightdark,Chowdhury:14darkmodes}.  
Similarly, structural symmetry breaking also allows for the emergence of Fano-like resonances\cite{BorisNatMaterialsFanoResonance, Cao:12ultrahighQdarkmode,Dong:10asymmetridarkmode}. 
The Fano spectral feature is usually defined as a resonant scattering phenomenon that gives rise to an asymmetric lineshape due to the interference between a broad spectral line and a narrow discrete resonance \cite{Fano61,Lukyanchuk10, Hao08}.  
Such a sharp resonance has been exploited for ultra-sensitive sensing and could lead to the design of narrow band THz emitters/detectors and highly selective filters. 
However, the high quality (Q)-factor of the Fano-like resonance is obtained at the expense of its intensity, which makes the Fano signature very challenging to be measured with low resolution and low signal-to-noise ratio systems.
Therefore, it is extremely important to excite high Q resonances with strong intensities and a Figure of Merit (FoM) can be studied to determine the optimal asymmetric parameter\cite{Cong15}.
%
%
%
%
%

In this paper, we report a systematic comparison of the FoM for asymmetric line shaped resonances in asymmetric planar THz metamaterials designed from a parent 4-gap SSR exhibiting four fold symmetry. 
First, we introduce asymmetry through shifting pairs of metallic arms (Technique 1) such that we achieve near-neighboring interactions between meta-atoms similar to our recent investigation of the 4-gap circle SSR\cite{Burrow17}. 
Second, we incorporate structural asymmetry by shifting the capacitive gaps (Technique 2) generating Fano-like resonances\cite{PhysRev.124.1866FANO}. 
Last, we adapt a new method of asymmetry (gap widening) to Technique 2 which doubles the FoM and increases the number of modes within a 300 GHz window. 
Numerical investigations and semi-analytical models are applied to provide further insights on each asymmetric case which are in good agreement with the experimental observations. 
The results of this work demonstrate an effective process for the design of THz metasurfaces for sensing and modulation applications where multi-mode, high FoM characteristics are desirable.

\section{DESIGN, FABRICATION AND EXPERIMENTAL CHARACTERIZATION OF SYMMETRICAL METAFILM}
We begin by describing the geometric parameters, numerical simulations and experimental techniques for the symmetric metafilm [Fig. 1(k)]. 
The relevant geometric parameters for the symmetric unit cell are the periodicities $P_{x} = P_{y} = 300 ~\mu\text{m}$, total length and width of the metallic ring $a = 250 ~\mu\text{m}$ and width of the ring $w = 35~\mu\text{m}$. Numerical calculations were carried out using the finite element method (FEM). 
In these calculations, the metafilm was illuminated at normal incidence, under TE-polarization ($E~//~y$-axis) or TM-polarized radiation ($E~//~x$-axis). 
Periodic boundary conditions were applied in the numerical model in order to mimic a 2D infinite structure. 
In simulations, the polyimide substrate was treated as a dielectric with $\varepsilon = 3.3 + i0.05$ \cite{Yahiaoui15,Yahiaoui18} and the silver (Ag) was modeled as a lossy metal with a conductivity of $3.1\times10^{7} \text{ S/m}$ to account for lower conductivities commonly observed in actuality. 

Each metafilm was fabricated using standard optical lithography to deposit 100 nm of Ag with a 10 nm adhesive layer of chromium on commercially available 50.8 $\mu$m polyimide substrates.  
Transmission measurements were performed using linearly polarized collimated radiation from a continuous-wave (CW) THz spectrometer (Teraview CW Spectra 400). The high spectral resolution (100 MHz) is maintained by the precision of the temperature tuning of two near-IR diode lasers and not by a mechanical delay stage as found in a conventional THz time-domain spectroscopy. 
The transmission spectrum from each  sample was determined as $T(\omega) = P_{M}(\omega)/P_{sub}(\omega)$, where $P_{M}(\omega)$ and $P_{sub}(\omega)$ are the filtered THz power spectra of the planar metafilm and flexible substrate respectively.

\begin{figure}[ht!]
%
%
%
\includegraphics[width=\linewidth]{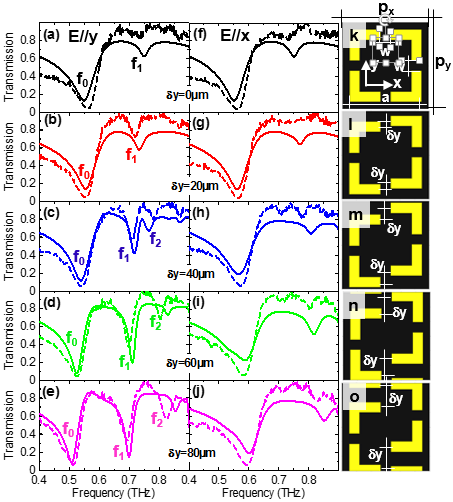}
\caption{Evolution of the simulated (solid lines) and measured (dashed lines) transmission spectra for an y-polarized (a)-(e) and x-polarized incident THz wave (f)-(j). (k)-(o) Schematics of the SRR unit cells with various degrees of asymmetry.}
\label{fig:1}
\end{figure}

The simulated (solid black line) and measured (dashed black line) transmission spectra for the symmetric 4-gap metafilm are plotted in Fig.~1(a) and are in good agreement with each other. 
A fundamental resonant feature $f_{0} \sim 0.56~\text{THz}$ and a weakly pronounced higher order transmission dip $f_{1} \sim 0.75~\text{THz}$ are observed with simulated Q-factors $Q_{0} = 5.94$ and $Q_{1} = 15.52$, respectively. 
Here, the Q-factor is defined as Q = $f_{0}/\Delta f$ where $f_{0}$ is the resonant frequency and $\Delta f$ is the full width at half maximum (FWHM) of each transmission dip. 
The parameters were extracted from a least-squares fit to an asymmetric lineshape profile of the transmission spectrum defined as

\begin{equation}
T(f) = A\frac{[\Lambda + (f - f_{0})/\Delta f]^{2}}{1 + (f - f_{0})^{2}}
\end{equation}
where $A$ is the amplitude, $f_{0}$ is the resonant frequency, $\Lambda$ yields the lineshape asymmetry parameter and $\Delta f$ is the full width half maximum (FWHM) of the mode \cite{Singh:11}.
%
%

Quality factor and resonance intensity (referred to as modulation depth (MD) herein and defined as the difference between the maximum and minimum transmission amplitude of the resonance) are two important parameters used to characterize modes for photonics and sensing applications; therefore, we consider FoM calculations of each transmission dip as $FoM = Q \times MD$ \cite{Cong15}. 
Due to the four-fold rotational symmetry imposed by the geometry of the design, the metafilm has identical response for an incident TM-polarized radiation (i.e., $E~//x~$-axis) [see Fig. 1(f)].

\section{TECHNIQUE 1: L-BRACKET SHIFTING ASYMMETRY}
\begin{figure}[ht!]
\includegraphics[width=\linewidth]{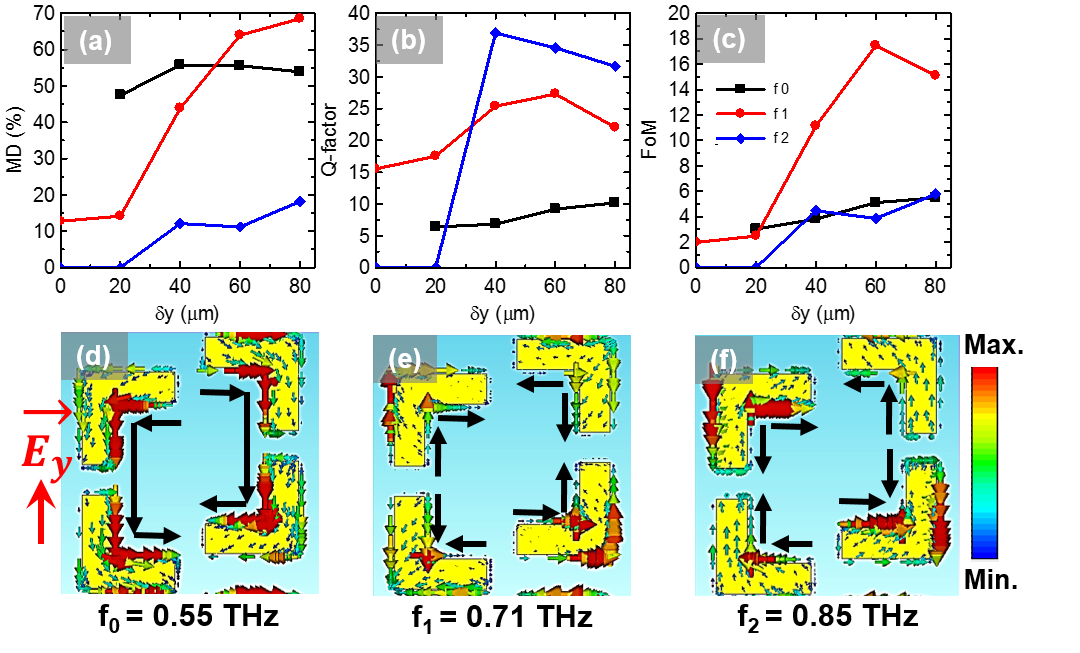}
\caption{(a) Modulation depth, (b) Q-factor and (c) Figure of merit (FoM) of $f_{0}$, $f_{1}$, and $f_{2}$ as a function of asymmetric shift. (d)-(f) Surface current distributions of asymmetric structure ($\delta y = 60~ \mu m$) at 0.55 THz, 0.71 THz and 0.85 THz, respectively}
\label{fig:2}
\end{figure}

The first methodology introduces asymmetry into the design by offsetting the right and left arms of the SRR a distance $\delta_{y}$ ranging between 0$\text{ }\mu\text{m}$ - 80$\text~\mu\text{m}$ along the $y$-axis, as shown in Figs. 1(l)-1(o), similar to our previous work\cite{Burrow17}. Wang \textit{et. al.} disassociated a single arm of a circle SSR via translation which resulted in an ultra-high Fano-like mode\cite{WANG201560}.  
In the case where $E~//~ y$-axis, as $\delta_{y}$ increases, $f_{0}$ and $f_{1}$ red shift and the modulation depth of $f_{0}$ remains relatively constant as $f_{1}$ increases, depicted in Fig. \ref{fig:2}(a). 
When $E~//~ x$-axis, we observe the excitation a similar broad fundamental resonance and a weakly coupled higher order mode that both blue shift as a function of asymmetry as depicted in Figs. 1(f)-1(j). 
Both modes in this polarization case exhibit low FoMs and are in turn neglected in mode analysis. 

With respect to the case where $E~//~ y$-axis, the redshifts of the two transmission dips, $f_{0}$ and $f_{1}$, to lower frequencies can be understood within an equivalent circuit scheme. 
For example, SRRs at THz frequencies are analogous to micro-scale LC resonators where effective inductance arises from the loop formed by the SRR and the effective capacitance is built up within the gap region between the SRR arms. 
Altering the structural geometry of the resonators results in a change in the effective inductance (L) and effective capacitance (C). 
Therefore, the frequency of the transmission dip will also be changed. 
Offsetting the right and left arms of the SRRs results in an increase of the equivalent length of the resonators, which suggests an increase in the equivalent inductance involved in the LC resonance. 
Since the resonance frequency $f_{LC}$ is inversely proportional to the square root of the inductance (L) given by the following relation: $f_{LC}=1/(2\pi\sqrt{LC})$, the spectral feature shifts to lower frequencies with increasing the equivalent inductance.

Additionally, with increasing  asymmetry, a high Q mode $f_{2}$ emerges that increases in modulation depth. 
Interestingly, the excitation of $f_{2}$ creates a transmission window similar to the results reported in the 4-gap circle SRR with the exception of the spectrally trapped mode $f_{1}$ between the fundamental and higher order mode \cite{Burrow17}. 

To better understand the origin of the resonances, we have plotted the surface current distributions at the frequency resonances $f_0$, $f_1$ and $f_2$ for $\delta = 60 \text{ }\mu\text{m}$), as shown in Figs. \ref{fig:2}(d)-\ref{fig:2}(f). For the fundamental mode $f_{0} = 0.55~\text{THz}$, the surface currents mainly flow in the right and left arms of the SRR, where the currents vertically converge to the bottom gap and diverge from the top gap, thus suggesting a dipole-like resonant behavior [Fig. 3(a)]. For the higher order resonance frequencies $f_{1} = 0.71 ~\text{THz}$ and $f_{2} = 0.85~\text{THz}$, the surface current flows are all with chaotic directions and with distinctive nodes, as shown in Figs. \ref{fig:2}(b)-\ref{fig:2}(c). In this case, the top and bottom horizontal arms of the SRR show antiparallel surface current distributions, as well as the vertical right and left arms. 

As previously mentioned, since the MD is enhanced at the cost of Q factor, it is useful to explore the trade-off of the Q-factor [see Fig. \ref{fig:2}(b)] and monotonically increasing modulation depth [see Fig. \ref{fig:2}(a)]\cite{Cong:15}. 
The FoMs for Method 1 is calculated and plotted in Fig. \ref{fig:2}(c) for each transmission feature. 
Although the higher order resonance $f_{2}$ exhibits a high Q-factor of 36.91 for $\delta y = 40~\mu m$ the incident THz field is weakly coupled resulting in a MD = 12\%, and further exhibiting a relatively low FoM when compared to other resonances. The maximal FoM value of 17.47 for $\delta = 60~\mu$m is observed for $\delta y = 60~\mu m$ where the SSR begins to couple to the neighboring unit cell of the metasurface. 
%
%

\section{TECHNIQUE 2: DUAL GAP TRANSLATION ASYMMETRY}

\begin{figure*}[ht!]
\includegraphics [width = 0.7\linewidth] {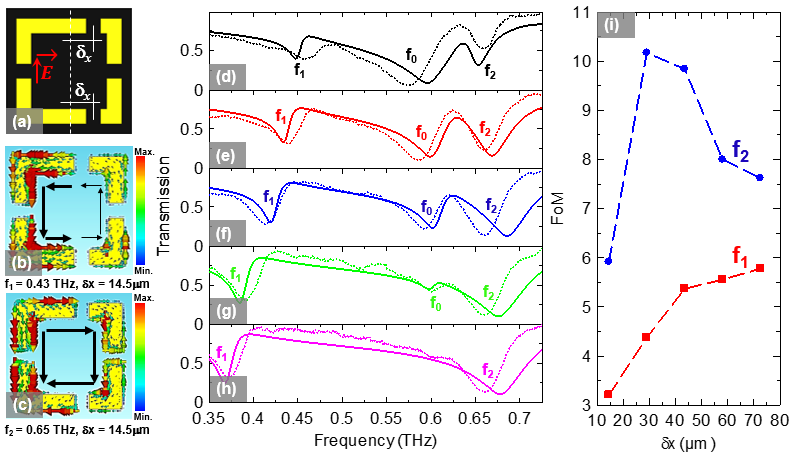}
%
%
%
\caption{(a) Unit cell of the designed asymmetric SRR MM. (b)-(c) distributions of surface current at two corresponding resonance frequencies $f_{1} = 0.43$ THz and $f_{2} = 0.65$ THz when $\delta_{x}$ = 14.5 $\mu$m. (d)-(h) Evolution of the simulated (solid lines) and measured (doted lines)  transmission spectra for different values of $\delta_{x}$. (i) Measured FoM curves for $f_{1} = 0.43$ THz and $f_{2} = 0.65$ THz, respectively.}
\label{fig:5}
\end{figure*}

Beginning with a fully symmetric 4-gap SRR array, the second asymmetry can be generated by simultaneously right-shifting the top and bottom capacitive air gaps a distance $\delta_{x}$ in the range 14.5 - 72.5 $\mu$m, as shown in Fig. \ref{fig:5}(a). This method of asymmetry has been extensively studied to exhibit an effective route to excite Fano-like resulting from the interference between a sharp resonance and a smooth continuum-like spectrum\cite{Cao2015TuningFanoTwoGap}. 
The simulated (solid lines) and measured (dashed lines) transmission spectra for increasing values of $\delta_{x}$ are shown in Figs. 3(d)-3(h). 
Despite of some deviation between the simulation and experimental results, which may be mainly due to imperfections in sample preparation, they are overall in quite good agreement with one another. 
%
%
%
%

One can clearly observe the excitation of two asymmetric line shaped modes at $f_{1} = 0.43$ THz and $f_{2} = 0.65$ THz, which represent Fano-like resonances with Q-factors of about 29.71 and 39.18, respectively. 
To understand the nature of the double Fano excitations, we have simulated in Figs. \ref{fig:5}(b) and (c) the surface current distributions when $\delta_{x}$ = 14.5 $\mu$m at $f_{1}$ and $f_{2}$, respectively. 
For the first sharp resonance at 0.43 THz, we observe anti-parallel currents in the right and left arms of the SRR. 
Anti-symmetric currents create fields that interfere destructively resulting in radiation suppression and low-loss light propagation. 
This particular resonance is similar in nature to the inductive capacitive (LC) resonance in a single-gap SSR, as the resonance results in current configuration that forms a closed loop, which gives rise to a magnetic dipole moment perpendicular to the MM plane.
The surface currents at $f_{2}$ = 0.65 THz flow in such a way that the unit cell of the structure can be geometrically regarded as the combination of an U-shaped resonator (USR) and a cut wire resonator (CWR).  
In this case, the resonance is due to a dipole-like parallel current distribution as seen in Fig. \ref{fig:5}(c).
Additionally, it can be seen that the intensity of $f_{1}$ increases with increasing the degree of asymmetry (i.e. increasing $\delta_{x}$) with a modulation depth of about 40$\%$ for $\delta_{x}$ = 14.5 $\mu$m and around 75$\%$ for $\delta_{x}$ = 72.5 $\mu$m. 
Similarly, for $f_{2}$ the modulation depth increases as $\delta x$ increases. 
%
%
%
%
%
%

The degree of asymmetry is an important parameter controlling the electromagnetic resonant frequency and transmitted intensity\cite{Cong2017DegreeofAsy}. According to Figs. 3(d)-3(h), one can observe that the degree of asymmetry also affects the line-width and thus the Q-factor of the Fano resonances. In the case of lowest degree of asymmetry (i.e. $\delta_{x}$= 72.5 $\mu$m), the high Q-factor asymmetric Fano resonance $f_{2}$ in the SRR arises from the structural asymmetry which leads to interference between sharp discrete resonance and a much broader continuum-like spectrum of dipole resonance $f_{0}$. This narrow resonance arises from a sub-radiant dark mode for which the radiation losses are attenuated due to the weak coupling of the structure to free space \cite{Singh:11}. Such dark modes are exploited to demonstrate EIT-like effects in MMs, which could opens up avenues for designing slow light devices with large group index \cite{YahiaouiPRB18}. The resulting close proximity between the fundamental and high frequency resonances creates a strong coupling effect that reshapes the transmission curves as shown in Figs. 3(d)-(h). The Fano-like resonance mode located at $f_{1}$, however, is far ahead from the broad dipole mode, therefore, coupling with $f_{0}$ will not take place. 
 
Upon increasing the asymmetry, the dipole resonance $f_{2}$ broadens and increases in intensity (decreases in transmission amplitude), since the SRR metafilm couples more efficiently to the free space and becomes highly radiative. Interestingly, the fundamental broad dipole-like resonant mode that initially appears at $f_{0}$ gradually decreases in intensity and is completely suppressed in the highest degree of asymmetry (i.e., $\delta_{x}$ = 72.5 $\mu$m), thus completing a switching between single-mode $f_{0}$ and dual-mode operations $f_{1}$ and $f_{2}$. This may paves the way towards the development of reconfigurable photonics devices based on symmetry-breaking planar THz MMs.  For an intermediate value of $\delta_{x}$ = 29 $\mu$m, one can observe a narrow transparency window at 0.62 THz, with an amplitude as high as 72$\%$ between two quasi-symmetric resonance dips at around $f_{0}$ = 0.58 THz and $f_{2}$ = 0.66 THz, respectively, which is very similar to an electromagnetically induced transparency (EIT) signature [see Fig. 3(e)].

Here again, since the Fano resonance intensity is enhanced at the expense of the Q-factor, it is crucial to explore the trade-off between the resonance intensity and the Q-factor by calculating the FoM. Figure 3(i) shows the calculated FoM for both Fano-like modes $f_{1}$, and $f_{2}$. As we can see, the best numerical FoM of 10.17 with Fano resonance intensity (modulation depth MD) of 0.49 and Q-factor of 20.59 is obtained for $f_{1}$, at which the asymmetry parameter $\delta_{x}$ = 29 $\mu$m.

\section{TECHNIQUE 3: INCREASING CAPACITIVE GAP VOLUME}

\begin{figure*}[ht!]
\includegraphics [width=\linewidth] {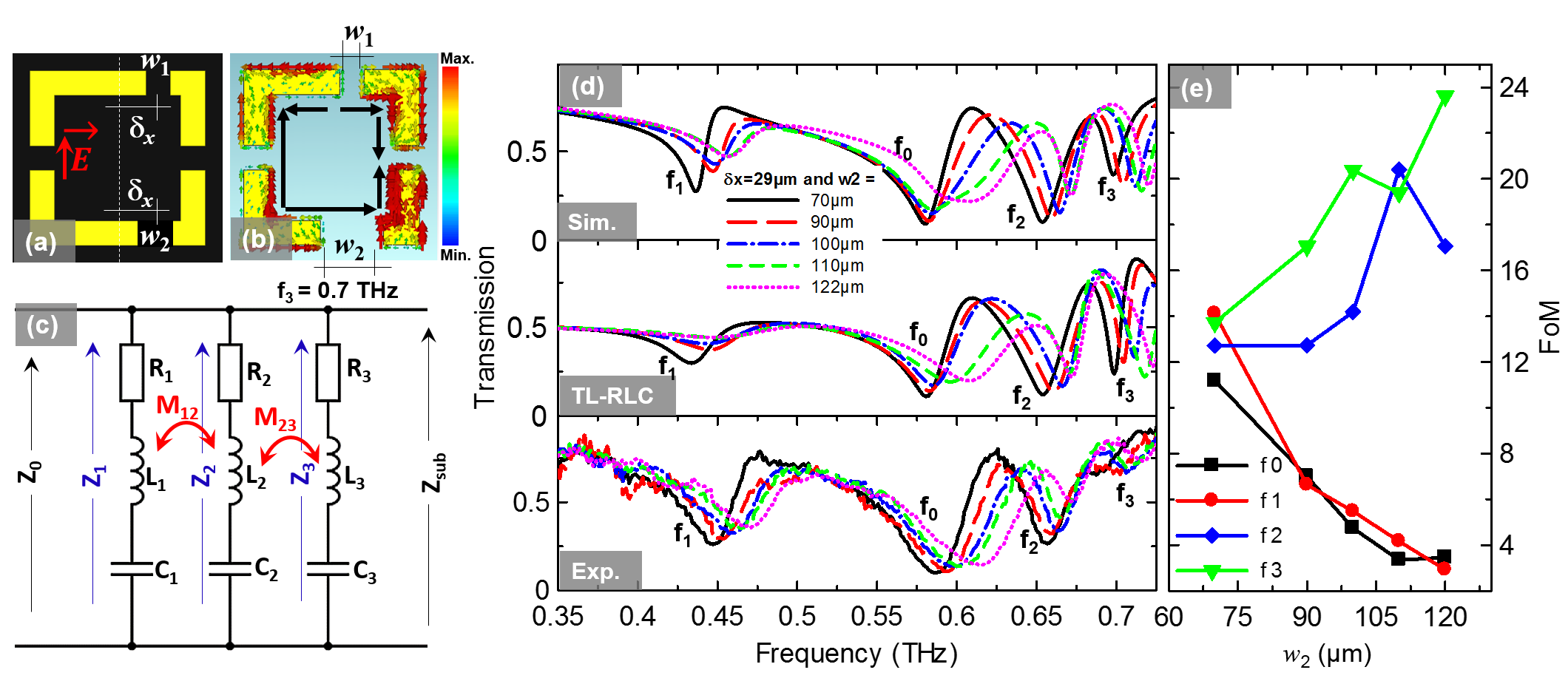}
\caption{(a) Schematic view of the designed wide gap asymmetric metasurface with $\delta_{x}$ = 29 $\mu$m,  $w_{1}$ = 35 $\mu$m and $w_{2}$ varying in the range 70 $\mu$m - 122 $\mu$m. (b) Surface current distribution in a single MM unit cell at $f_{3} = 0.7 \text{ THz}$, when $\delta_{x}$ = 29 $\mu$m,  $w_{1}$ = 35 $\mu$m and $w_{2}$ = 100 $\mu$m. (c) Circuit model for metafilm samples. (d) Simulated, analytical and measured transmission spectra for $\delta_{x}$ = 29 $\mu$m,  $w_{1}$ = 35 $\mu$m and $w_{2}$ varying in the range 70 $\mu$m - 122 $\mu$m. (e) Simulated FoM curves for $f_{0}$, $f_{1}$, $f_{2}$ and $f_{3}$, respectively.
\label{fig:widegap.}}
\end{figure*}

In order to increase the number of resonances and FoMs stemming from the optimal design in the previous case, let us consider the asymmetric structure investigated in Fig. 3(a) with an intermediate value of $\delta_{x}$ = 29 $\mu$m, where multiple modes with high intensity are obtained [see Fig. 3(e)]. By modifying the width of the bottom capacitive gap $w_{2}$  ($w_{1}$ is kept unchanged as well as the other geometrical parameters), another asymmetry will be created, as shown in Fig. 4(a). Figure 4(d) show the simulated, analytically modeled and measured transmission spectra through the metasurface for different values of $w_{2}$ with fairly good agreement. As can be seen, an additional Fano resonance emerges at $f_{3}$ = 0.7 THz. The surface current flows at 0.7 THz are all with chaotic directions and with distinctive nodes, as shown in Fig. 4(b).

By gradually increasing the width of the bottom capacitive gap $w_{2}$, one can observe a blueshift of the transmission spectrum in both simulations and measurements. This noticeable blueshift of the spectral features can be interpreted in the context of an equivalent circuit model. In general, the fundamental resonance frequency of SRRs-based metasurfaces is approximately given by $f_{0}=1/(2\pi\sqrt{LC})$, where L is the loop inductance and C is the gap capacitance. At resonance, the incident electric field induces a large accumulation of surface charges in the metal strips forming the gaps, causing a strong electric field confinement in the capacitive gaps. When the structural geometry of the SRR MM is altered by increasing the width of $w_{2}$, there is a decrease in the effective capacitance. As the resonance frequency is inversely proportional to the capacitance, the resonance frequency shifts to higher frequencies with increasing the gap width.

In order to confirm the resonant behavior of the 4-gap asymmetric SRR metasurface, we employ a semi-analytic transmission line (TL)-RLC model \cite{Ohara07}. The circuit model of our design under the TL theory is shown in Fig. 4(c), which is represented by three RLC circuits in parallel combination and inductively coupled through mutual inductances M12 and M23, respectively. We assumed SRR as an equivalent RLC circuit where it is typically considered that split gap corresponds to the capacitive part, the SRR loop corresponds to the inductive part and the internal reactance of SRR is represented by the resistance part. From the circuit model shown in Fig. 3(c), one can calculate the circuit impedance $Z_{ct}(\omega)$ as,
\begin{equation}
{Z}_{ct} (\omega)=\frac{ Z_{1,2} Z_{3} + \omega^{2}M^{2}}{ [Z_{1,2}+Z_{3} -2j\omega M]}
\end{equation}

\noindent where $\omega$ and $M$ represent the angular frequency and the mutual inductance respectively. $Z_{1,2}$ is the equivalent impedance due to the parallel combination of  circuits $R_{1}L_{1}C_{1}$ and $R_{2}L_{2}C_{2}$, respectively.  $Z_{3}$  corresponds to the impedance of circuit $R_{3}L_{3}C_{3}$. These complex impedances can be written as ${Z}_{i}(\omega)= {R}_{i} + j(\omega L_{i} - 1 /\omega C_{i}$), where $i$ = 1, 2 and 3, respectively. The equivalent impedance $Z_{1,2}$ is given by
\begin{equation}
{Z}_{1,2} (\omega)=\frac{ Z_{1} Z_{2} + \omega^{2}M_{1,2}^{2}}{ [Z_{1}+Z_{2} -2j\omega M_{1,2}]}\text{.}
\end{equation} 

One can note that the circuit impedance $Z_{ct}$ does not include the impedance due to the dielectric substrate. In Fig. 4(c), $Z_{0}$ and $Z_{sub}$ represent impedances of free space and polyimide substrate, respectively. The values of $Z_{0}$ and $Z_{sub}$ are 377 $\Omega$ and 219.5 $\Omega$ ($Z_{0}/\sqrt{\epsilon_{r}})$, where $\epsilon_{r}$ is the relative dielectric constant of the polyimide film), respectively. The overall impedance $Z(\omega)$ of our design including the effect of $Z_{ct}$ and $Z_{sub}$ in parallel combination can be written as

\begin{equation}
{Z}(\omega)=\frac{ Z_{sub}(Z_{1,2} Z_{3} + \omega^{2}M^{2})}{ Z_{sub}(Z_{1,2}+Z_{3} -2j\omega M) + (Z_{1,2} Z_{3})+ \omega^{2}M^{2})}\text{.}
\end{equation}

The normalized transmission amplitude, $T(\omega)$ of this transmission line-RLC circuit model is given by
\begin{equation}
{T}(\omega)=\frac{ 2Z(\omega)}{ Z_{0}+ Z(\omega)}\text{.}
\label{eqn:transamp}
\end{equation}

\begin{table*}[ht!]
\centering
\begin{tabular}{|c|c|c|c|c|c|c|c|c|c|c|c|c|c|c|c|}
	\hline
$g_{2}$ & $R_{1}$ & $ L_{1}$ & $ C_{1}$ & $R_{2}$ & $ L_{2}$ & $ C_{2}$ & $R_{3}$ & $ L_{3}$ & $ C_{3}$ & $R_{4}$ & $ L_{4}$ & $ C_{4}$ & $M_{1,2}$ & $ M_{1,2,3}$ & $ M_{1,2,3,4}$\\

($\mu$m) & $(\Omega)$ & (pH) & (fF) & $(\Omega)$ & (pH) & (fF) & $(\Omega)$ & (pH) & (fF) & $(\Omega)$ & (pH) & (fF) & (pH) & (pH) & (pH)\\

	\hline\hline
	70 & 30 & 180 & 0.7 & 5 & 235 & 0.34 & 5 & 222 & 0.265 &  0.5 & 180 & 0.28 &  -40 & -10.5& -3\\   
	\hline
	90 & 50 & 180 & 0.65 & 5 & 235 & 0.34 & 5 & 221 & 0.26 &  0.5 & 180 & 0.275 &  -40 & -10.5& -3\\   

	\hline
	100 & 70 & 180 & 0.65 & 5 & 235 & 0.335 & 5 & 221 & 0.255 &  0.5 & 180 & 0.267 &  -40 & -10.5& -6\\   
	\hline
	110 & 95 & 180 & 0.6 & 5 & 236 & 0.335 & 10 & 221 & 0.25 &  0.5 & 180 & 0.265 &  -45 & -3& -8\\   
	\hline
120 & 120 & 180 & 0.61 & 5 & 236 & 0.318 & 10 & 221 & 0.25 &  0.5 & 180 & 0.26 &  -45 & -2& -6\\   
\hline
	\end{tabular} 
\caption{Circuit parameters for matching TL-RLC model to simulated data. Resistor values are given in Ohms, capacitor values in femtofarads, and inductor values and coupling coefficients in picohenries.}
\label{t1}
\end{table*}

We used equation (\ref{eqn:transamp}) to calculate the transmission spectra and predict the resonant frequencies for specific values of $R_{1}, L_{1}, C_{1}, R_{2}, L_{2}, C_{2}, R_{3}, L_{3}, C_{3},  M_{1,2}$ and $M$, respectively, which are defined in Table 1. These values are obtained by fitting the transmission amplitude from the simulation using equation 4. One can notice that the calculated transmittance is in good agreement with the numerical simulations and experimental measurements.

The overall performance of the proposed series of devices shows a superior behavior to earlier designs. This proposed method allows for the excitation of four distinct modes over a 300 GHz window, which increases the potential for multi-spectral sensing applications. More importantly, the higher order modes exhibit high FoMs when compared to other methodologies, while exhibiting the Fano-like modes $f_{1}$ and $f_{2}$. 

\section{CONCLUSION}
To conclude, we have performed a resonance engineering investigation of  uniquely designed asymmetric metafilm devices, both numerically and experimentally, that exhibit multiple sharp Fano-like resonances in the THz frequency regime. The double asymmetric case out performs cases 1 and 2 in terms on number of excited modes and largest observed FoM when $\delta x = 29~\mu m$ and $w_{2} = 122~\mu m$. Additionally, the FoM for modes $f_{2}$ and $f_{3}$ exhibited a proportional relationship with asymmetry. A semi-theoretical circuit model was employed to explain the coupling between the incident THz field to the plasmonic modes of the structure.  Numerical simulations allow further interpretation of the experimental results obtained from a linearly polarized CW THz measurement system used to measure the transmission spectra of the proposed metadevices, which agree reasonably well. Our results are very promising suggesting optimized designs for THz driven switches, multi-spectral filters, and highly sensitive sensors.

\begin{acknowledgments}
We would like to acknowledge support the NASA Ohio Space Grant (NNX15AL50H) and the Air Force Office of Scientific Research (FA9550-16-1-0346). This work was performed in part at the Center for Nanoscale Systems (CNS), a member of the National Nanotechnology Coordinated Infrastructure Network (NNCI), which is supported by the National Science Foundation under NSF award no. 1541959.  J. A. B. acknowledges support from the Air Force Research Laboratory. T. A. S. acknowledges support from the CNS Scholars Program and R. Y. acknowledges support from the  W. M. Keck Foundation.
\end{acknowledgments}

\bibliographystyle{aipnum4-1}
\bibliography{references}

\end{document}